\documentclass{article}
\usepackage{algorithm} 
\usepackage{algpseudocode} 
\usepackage{url,hyperref,lineno,microtype,graphicx}
\usepackage[onehalfspacing]{setspace}

\begin{document}
\onecolumn

\title{HUXt - An open source, computationally efficient reduced-physics solar wind model, written in Python} 

\author{Luke Barnard and Mathew Owens \\ Department of Meteorology, University of Reading}

\maketitle

\section{Abstract}
HUXt is an open source numerical model of the solar wind written in Python. It is based on the solution of the 1D inviscid Burger's equation. This reduced-physics approach produces solar wind flow simulations that closely emulate the flow produced by 3-D magnetohydrodynamic (MHD) solar wind models at a small fraction of the computational expense. While not intended as a replacement for 3-D MHD, the simplicity and computational efficiency of HUXt offers several key advantages that enable experiments and the use of techniques that would otherwise be cost prohibitive. For example, large ensembles of $10^{2}-10^{5}$ members can easily be run with modest computing resources, which are useful for exploring and quantifying the uncertainty in space weather predictions, as well as for the application of some data assimilation methods.

In this article we present the developments in the latest version of HUXt, v4.0, and discuss our plans for future developments and applications of the model. The three key developments in v4.0 are: (1) a restructuring of the models solver to enable fully time-dependent boundary conditions, such that HUXt can in principle be initialised with in-situ observations from any of the fleet of heliospheric monitors; (2) new functionality to trace streaklines through the HUXt flow solutions, which can be used to track features such as the Heliospheric Current Sheet; (3) introduction of a small test-suite so that we can better ensure the reliability and reproducibility of HUXt simulations for all users across future versions. Other more minor developments are discussed in the article. 

Future applications of HUXt are discussed, including the development of both sequential and variational data assimilation schemes for assimilation of both remote sensing and in-situ plasma measures. Finally, we briefly discuss the progress of transitioning HUXt into an operational model at the UK's Met Office Space Weather Operations Center as part of the UK governments SWIMMR programme.

\section{Introduction}

Variability in the near-Earth space environment gives rise to space weather, which can have adverse effects on space- and ground-based technologies \cite{Cannon2013}. The largest disturbances to the terrestrial magnetospheric system are the result of coronal mass ejections (CMEs) arriving in near-Earth space \cite{gosling1993}. Thus, advanced forecasting of the arrival time and properties of CMEs at Earth is highly desirable \cite{riley2018}. While near-Sun CME properties -- notably speed and direction -- can be estimated from coronagraph observations \cite{gopalswamy2009}, forecasting at Earth also requires accurate knowledge of the variable solar wind conditions through which CMEs propagate  \cite{case2008}. 

The dynamic interaction between a CME and the ambient solar wind is typically modelled using time-dependent magnetohydrodynamic (MHD) simulations of the solar wind (e.g. \cite{merkin2016,Riley2001,Odstrcil2003, narechania2021}). These models capture the large-scale MHD fluid behaviour which governs much of the physics of CME propagation and evolution to Earth, and are a vital tool that enable both heliophysics research and space weather forecasting \cite{Mays2015,riley2018,odstrcil2020}.  More recently, it has been demonstrated that similar forecasting results, at least to first order, can be obtained with greatly simplified models \cite{riley2011, owens2020a}. Such models are not intended to replace the more sophisticated simulations, but open up complementary capabilities via greatly reduced computational overhead.

This paper is intended to briefly review the reduced-physics model, HUXt (heliospheric upwind extrapolation with time-dependence), and then highlight some of the functionality of the HUXt Python implementation.

HUXt has already proved useful in a number of contexts. \cite{barnard2020} demonstrated how an ensemble of HUXt runs could be weighted by comparison with STEREO Heliospheric Imager (HI) time-elongation profiles of CME fronts to return improved ensemble hindcasts of CME arrival times. \cite{chi2021} used HUXt simulations to examine the evolving structure of two interacting CMEs, finding that the HUXt simulations were consistent with the STEREO HI observations of these CMEs.

\cite{barnard2021} used HUXt simulations of Cone CMEs to examine the validity of the assumptions of CME geometric models and the ability of these models to reconstruct a CMEs kinematic profile from time-elongation profiles of a CMEs flank, such as those derived from HI observations. Similarly, \cite{hinterreiter2021a} used HUXt simulations as part of their work to introduce time-dependent structure to their ElEvoHI geometric model. These works both highlighted the importance of including time-dependent structure and solar-wind CME interactions for increasing the real-world representivity of CME modelling.

Additionally, separate from these CME focused studies, \cite{macneil2022} demonstrated how HUXt may be used to backmap in-situ solar wind observation to the source regions in the low corona, which is an important technique for estimating the origins of solar wind plasma structures. Furthermore, \cite{bunting2022} used HUXt simulations in their calibration and long-term validation experiments into using coronal tomography to generate inner boundary conditions for solar wind numerical models.

We now proceed to review both the background and the current status of our Python implementation of HUXt. Section \ref{sec:huxt_equations} describes the theoretical background to HUXt, whilst section \ref{sec:numerical_scheme} discusses the numerical scheme used to solve the discretised model equations. Section \ref{sec:testing} presents some new results of testing the performance of the numerical scheme. Section \ref{sec:functionality} describes some of they key functionality included in our Python implementation of HUXt. Finally, some developing applications of HUXt are described in Section \ref{sec:Summary}.

\section{The HUXt reduced-physics solar wind model}
\label{sec:huxt_equations}
HUXt takes a reduced-physics approach to modelling the solar wind flow, employing approximations to greatly reduce the complexity of the MHD momentum equation. A full derivation is provided in \cite{owens2020a} and so we provide only a summary here. Magnetic, gravitational, and pressure gradient forces are neglected, as in the solar wind these terms are typically small compared to the flow momentum \cite{riley2011}. Additionally, the solar wind is assumed to be purely radial, and so non radial terms are neglected. In this limit, the solar wind is modelled by the inviscid Burger's equation, 

\begin{equation}
\label{eq:burger}
\frac{\partial V_{r}}{\partial t} + V_{r}\frac{\partial V_{r}}{\partial r} = 0,
\end{equation}

where $V_{r}$ is the radial solar wind speed. \cite{pizzo1978} and \cite{riley2011} outlined this simplified description of the solar wind and made the further assumption of time-stationary flows to give the Heliospheric Upwind eXtrapolation (HUX) model \cite{reiss2020}. HUXt, however, maintains explicit time dependence, allowing structures such as coronal mass ejections (CMEs) and time-dependent ambient solar wind flow to be modelled.

The solar wind continues to accelerate throughout the inner heliosphere and in HUXt this residual acceleration, $\Delta V(r)$, is represented by an empirical parameterisation, which for a uniform solar wind is represented as  

\begin{equation}
\label{eq:vsw_eqn}
V_{r}(r) = V_{0} + \Delta V(r),
\end{equation}

where $V_{0}$ is the speed at a reference height and 
\begin{equation}
\label{eq:acc_term}
\Delta V(r) = \alpha V_{0}\left[ 1 - e^{\left(\frac{(r_{0} - r)}{r_{h}}\right)}\right],
\end{equation}

where $\alpha$ is the acceleration factor, set to be $0.15$, and $r_{h}$ is the scale height over which it applies, set to be $50~r_{\odot}$, whilst $r_{0}$ is the radial distance at the reference height corresponding to $V_{0}$, taken to be $30~r_{\odot}$. This form is designed to mimic that arising from the energy equations in MHD solar wind models \cite{riley2011}.

\cite{owens2020a} analysed the performance of HUXt relative to the HelioMAS MHD model of the solar wind, comparing each models representation of phenomena such as CIRs and Cone CMEs. Figure 4 of \cite{owens2020a} shows that HUXt and HelioMAS both return very similar CIR structures. However, there are small but systematic discrepancies. For example, in HUXt, CIRs tend to have sharper gradients and higher maximum speeds relative to HelioMAS. Two factors likely affect this; firstly, this could be due to non-radial dynamics in HelioMAS allowing flow deflections that HUXt does not model; secondly, the upwind numerical scheme used in HUXt is known to permit sharper gradients than the numerical schemes employed by MHD models. For a more complete review of the scientific justification of HUXt, and its comparison to MHD simulations, see \cite{owens2020a}.

\section{Numerical scheme}
\label{sec:numerical_scheme}

The HUXt model equations are solved numerically using a first order upwind scheme \cite{press2007}, and the discretised equations are given in \cite{owens2020a}.

The default radial grid of HUXt has an inner boundary at $30~r_{\odot}$, outer boundary at $240~r_{\odot}$, with a radial grid step of $1.5~r_{\odot}$. However, HUXt has built-in functionality to work with different radial grids and boundary heights, as discussed in Sections \ref{sec:backmap} and \ref{sec:insitubc}.

Fundamentally, HUXt is a 1-D radial model of the solar wind. Single 1D solutions are useful in their own right, due to the rapid computation time (much less than a second on a modest CPU for a 5-day simulation out to Earth orbit). Thus, one use-case is to run HUXt in a synodic frame of reference and simulate the Earth-Sun line. This can be done in large ensembles of O($10^3-10^5$), enabling case-specific estimates of forecast uncertainty \cite{owens2017c}. Examples of idealised and data-driven 1-D solutions are provided in the \textit{HUXt\_examples.ipynb} notebook, discussed in Section \ref{sec:python}. Here, the examples shown are for the more generalised case of 2-D and 3-D solutions that incorporate a range of longitudes and/or latitudes by assembling a set of the 1-D radial HUXt solutions. Such 2-D and 3-D solutions would generally be performed in the sidereal reference frame. 

The default longitudinal grid spans $0$ to $2\pi$ in 128 bins.  The default latitudinal grid has $45$ equally spaced cells in sine latitude. On initialisation, a single longitude or range of longitudes must be specified, whilst, unless otherwise specified, the latitude is assumed to be $0^{\circ}$. Finer or coarser grids can be used as required. The model grid sets zero longitude to that of Earth at the time of the start of the run. In the synodic reference frame, this is maintained throughout the run. For the sidereal reference frame, Earth increases in longitude by around $1^\circ$ per day. Note that because the 2-D and 3-D solutions are collections of independent 1-D radial solutions, any subset of the longitude and latitude grid can be considered, without any impact on the solution due to longitude or latitude boundary conditions. (Figure \ref{fig:huxt3d} shows an example of this functionality for a 3-D solution, discussed in Section \ref{sec:3d}.)

The model time-step, $\Delta t$ is set by the Courant–Friedrichs–Lewy (CFL) condition, $\Delta t \leq \frac{\Delta r}{v_{max}}$, with the default value of $v_{max}=2000~km~s^{-1}$ resulting in $\Delta t=8.70$ minutes. The default $v_{max}$ is chosen as a compromise that is suitable for all plausible ambient solar winds and the large majority of plausible Cone CME scenarios, whilst also not being set so high to be an inefficient use of computing resources.

To initialise HUXt, a uniform solar wind speed of $400~km~s^{-1}$ is set at all model coordinates. The model is then spun up for a time period that depends on the minimum solar wind speed in the time-dependent boundary conditions. The spin up time is set to 1.5 times the travel time it would take the slowest speed to traverse the model domain. The results of the spin up are discarded, meaning the user is presented with `usable' solar wind conditions from the start of the requested simulation time.

\subsection{Testing the numerical scheme}
\label{sec:testing}

HUXt was primarily developed as a surrogate for 3-D MHD simulations for situations where 3-D MHD simulations would be too computationally expensive or complex. Consequently, \cite{owens2020a} presented a thorough comparison between HUXt and HelioMAS simulations. Over a 40+ year period of 578 Carrington rotations, the mean absolute error in the ambient solar wind solutions of HUXt and HelioMAS was $25.6~km~s^{-1}$ or $6.4\%$. In this sense, the HUXt numerical scheme and default parameters were considered fit-for-purpose in serving as an surrogate for 3-D MHD solar wind simulations. Here we present new analysis that examines the consistency and convergence of the discretised HUXt numerical scheme in approximating the continuous solutions to the model equations.

\begin{figure}[ht]
\begin{center}
\includegraphics[width=15cm]{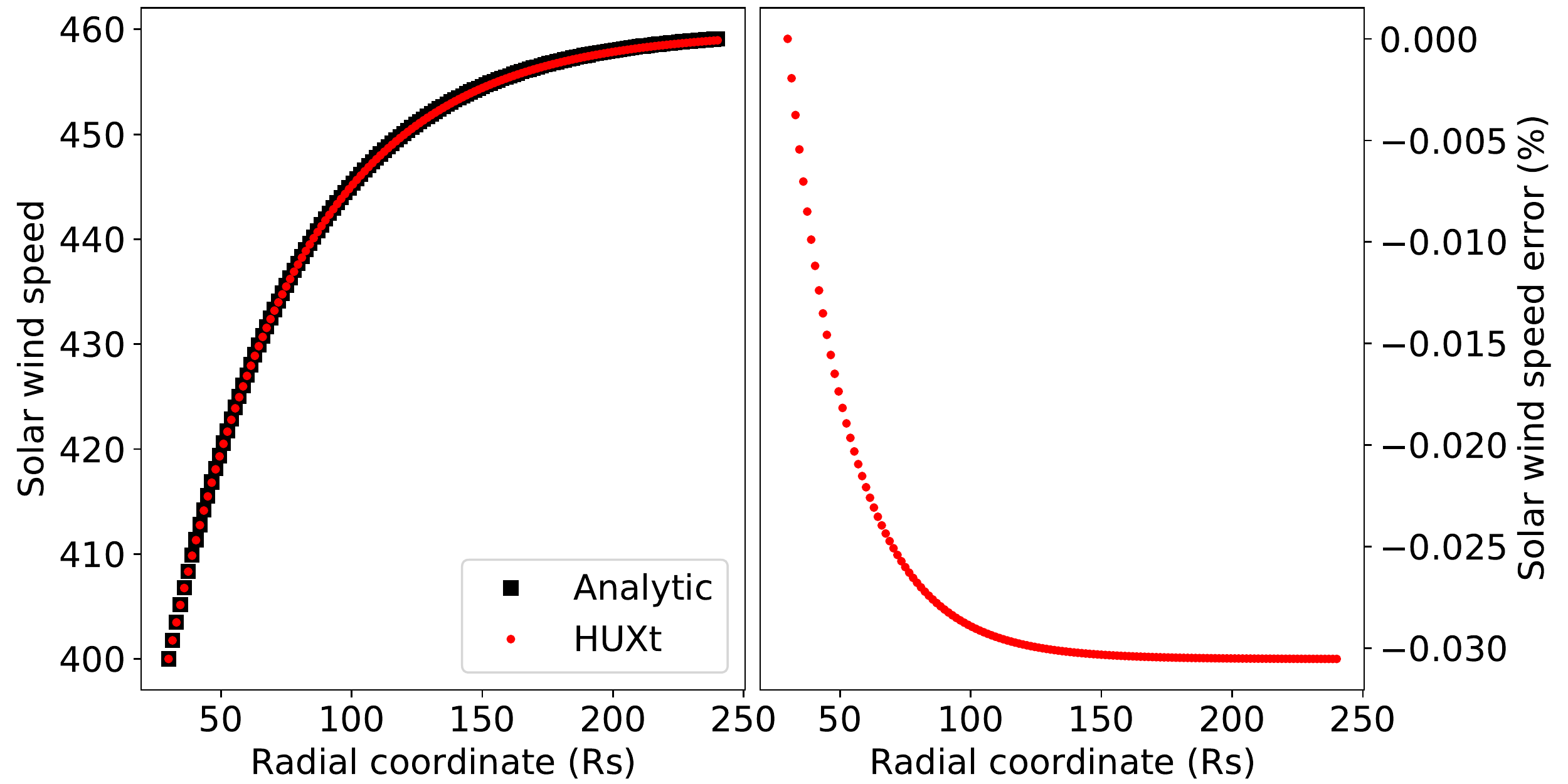}
\end{center}
\caption{Comparison of the analytic and numerical solutions to the HUXt equations for a simplified scenario of uniform and stationary inner boundary conditions. (Left) The solar wind speed as a function of radius for the analytic (black) and HUXt solutions (red). (Right) The percentage error of the HUXt solution relative to the analytical solution as a function of radius.}\label{fig:analytical_error}
\end{figure}

To assess the consistency of HUXt with a solution to the continuous model equations, we first compare the numerical solution with the analytical solution for the simple scenario of a uniform and stationary inner boundary condition. Figure \ref{fig:analytical_error} compares the solutions for a constant inner boundary condition of $400~km~s^{-1}$. There is excellent agreement between the HUXt and analytical solution, with a very small negative bias that approaches an asymptotic value of approximately $-0.03\%$. Further experiments show that the magnitude of this error is a function of the radial grid step and so we conclude that this error is related to the discretisation of the HUXt residual acceleration equation. However, the magnitude of this error is insignificant compared to the uncertainties relating to observationally derived boundary conditions (e.g. from coronal models) and from the simplifying physical assumptions used to derive the HUXt equations.

The second test is convergence testing, which seeks to assess if and how the numerical solutions are converging towards an exact continuous solution as the discretised grid steps are reduced in size. We ran HUXt for a time dependent scenario of a bimodal solar wind, with slow and fast streams of $350~km~s^{-1}$ and $600~km~s^{-1}$ at the inner boundary. The model ran for 13.5 days, along a single longitude. The simulations of this scenario were generated using a range of decreasing radial grid steps, with $dr$ being in the set $\{12, 6, 3, 1.5, 0.75, 0.375, 0.188, 0.094, 0.047\}$, in units of $R_{\odot}$. The finest radial grid step, $dr=0.047R_{\odot}$, was used as a reference standard to compare the other grid steps against.

\begin{figure}[ht]
\begin{center}
\includegraphics[width=15cm]{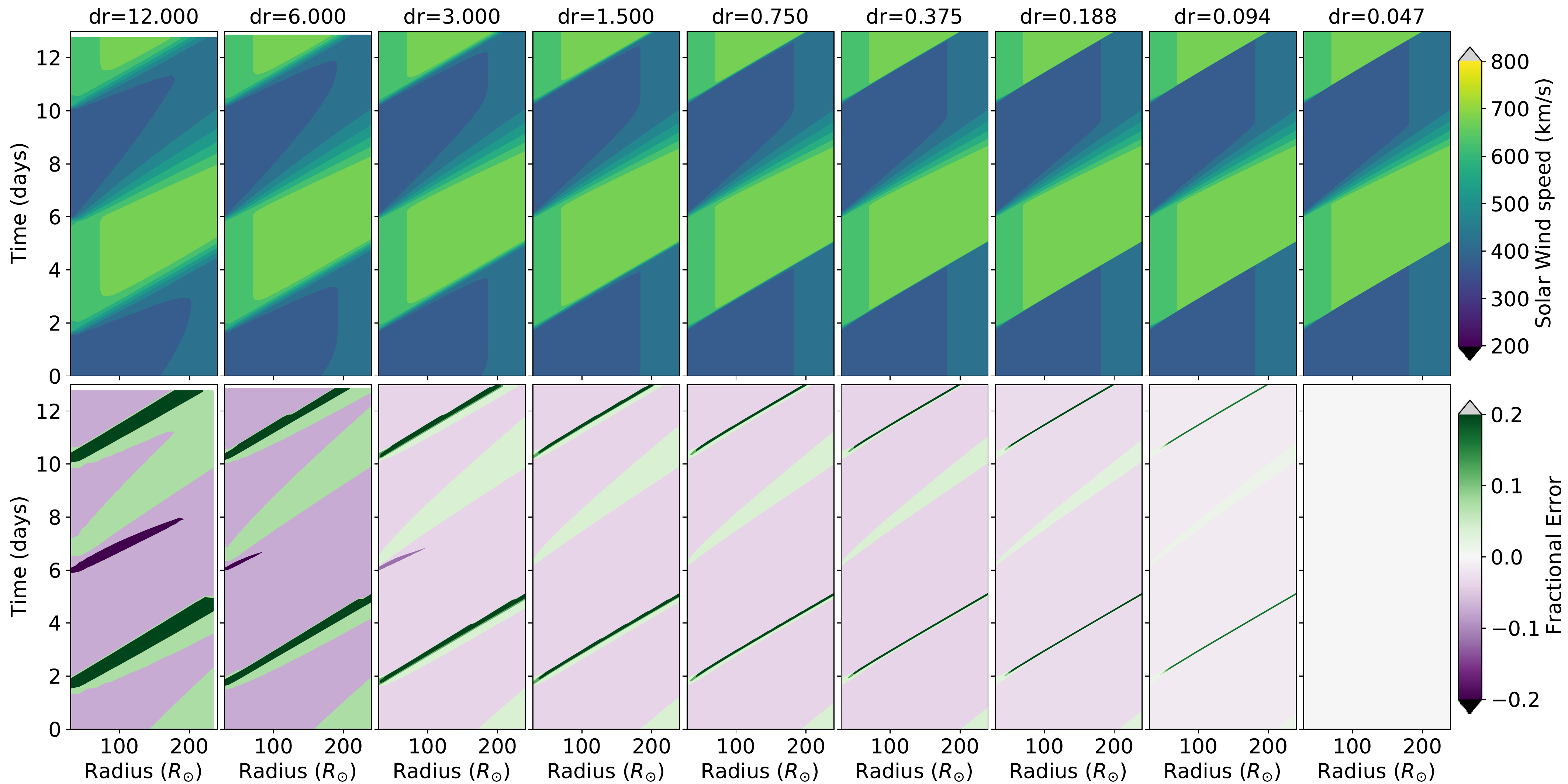}
\end{center}
\caption{(Top) Hovm\"{o}ller (distance-time) diagrams of the HUXt solar wind speed solutions for the set of different radial grid steps, with radial grid step decreasing from right to left. (Bottom) Hovm\"{o}ller diagrams of the fractional error in the HUXt solution at a given radial grid step relative the HUXt solution at the finest radial grid step of $dr=0.047~R_{\odot}$. }\label{fig:convergence_test}
\end{figure}

Figure \ref{fig:convergence_test} presents the results of these simulations. For the largest values of $dr$, there are large regions of significant disagreement with the reference standard. But these errors rapidly decrease with decreasing grid step, which indicates that the model is converging. As the grid step decreases the errors become more localised to the shock regions between fast and slow streams. 

\begin{figure}[ht]
\begin{center}
\includegraphics[width=15cm]{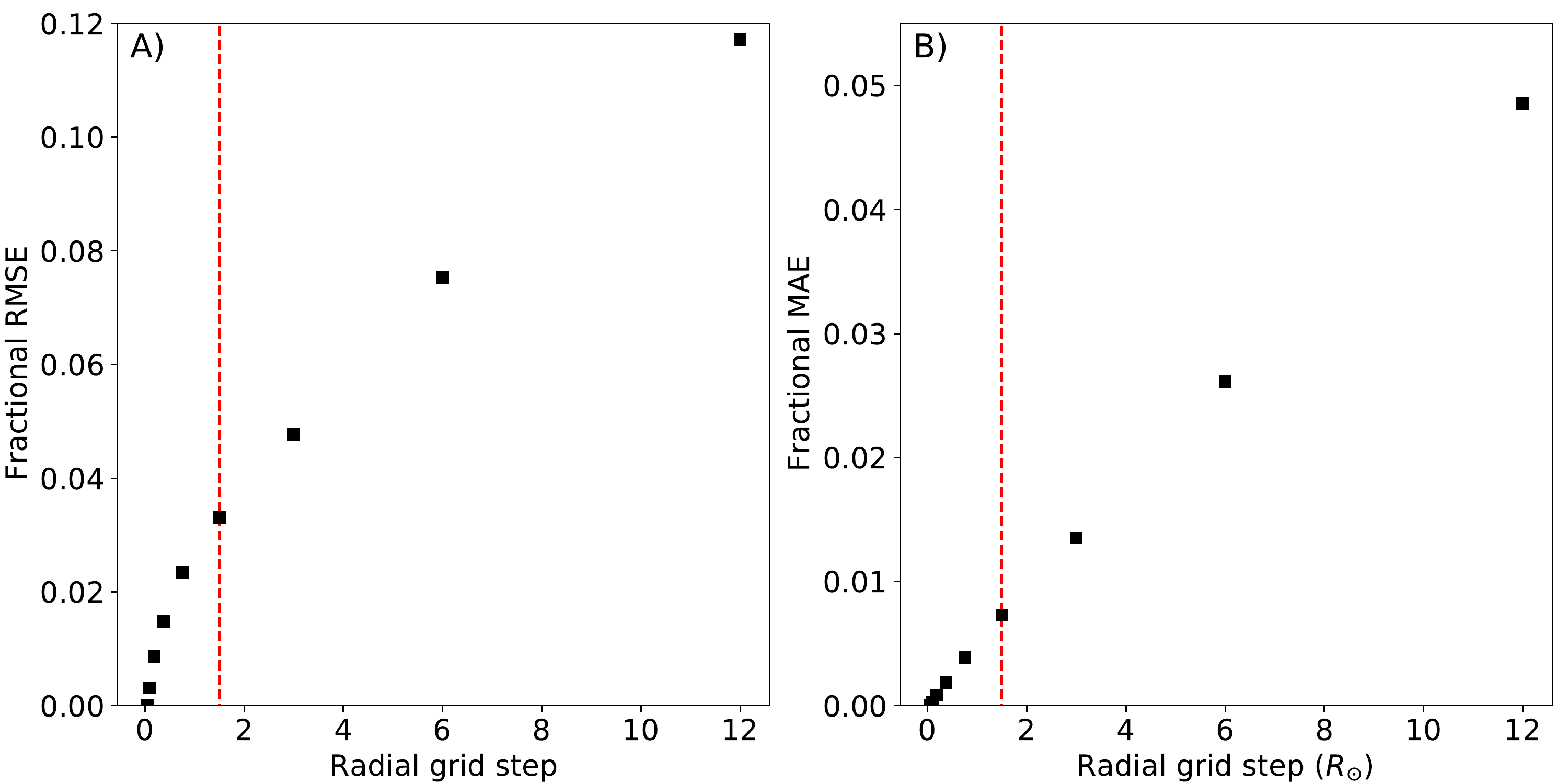}
\end{center}
\caption{(A) The fractional root mean square error (RMSE) of the convergence tests in Figure \ref{fig:convergence_test} as a function of radial grid step. (B) The corresponding fractional mean absolute error (MAE) as a function of radial grid step. The red dashed lines mark the default HUXt radial grid step of $1.5~R_{\odot}$.}\label{fig:dr_vs_error}
\end{figure}

Assessment of the domain over which the model can be said to have converged is subjective, and depends on the acceptable tolerance of error for a particular application. For each of these radial grid steps, we computed both the mean absolute error (MAE) and root mean square error (RMSE). Figure \ref{fig:dr_vs_error} presents the fractional MAE and RMSE data as a function of radial grid step. This summarises what could be inferred from Figure \ref{fig:convergence_test}, that the errors do decrease with grid step. The red dashed line marks the $1.5~R_{\odot}$ grid step, which is the default configuration for HUXt. At this radial grid step, the errors are modest, with MAE being $0.7\%$ and RMSE being $3.3\%$. Therefore, we consider $1.5~R_{\odot}$ to be a sensible upper limit on the radial grid step, which was chosen as the default $dr$ to balance the requirement that HUXt run efficiently against the discretisation errors associated with larger radial grid steps. It is simple to specify other radial grid steps in HUXt by modifying the \textbf{huxt\_constants} function in \emph{huxt.py}.

\section{Python implementation}
\label{sec:python}
\begin{figure}[ht]
\begin{center}
\includegraphics[width=15cm]{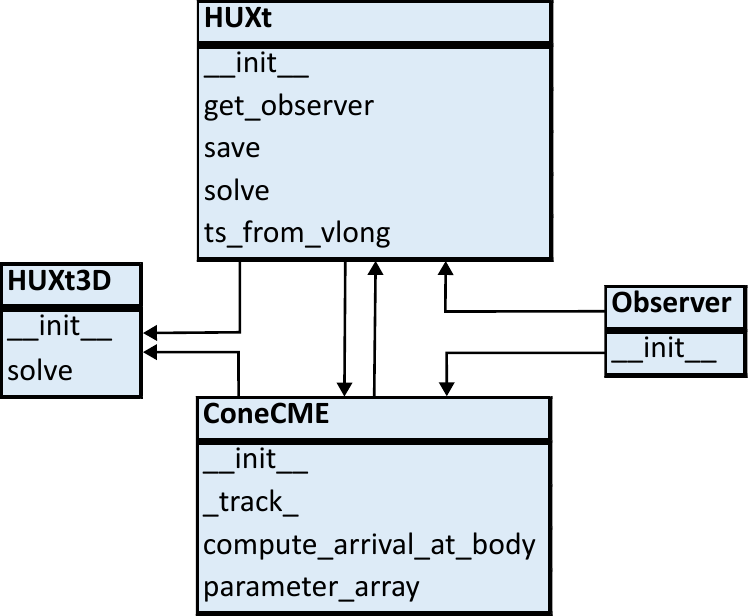}
\end{center}
\caption{This diagram highlights the classes used within HUXt, and their associated methods, and the relationships between them.}\label{fig:huxt_diagram}
\end{figure}

A Jupyter notebook of examples, \textit{HUXt\_examples.ipynb}, is provided with HUXt, which shows in detail how HUXt can be used in different scenarios. Here we provide a brief overview of this Python implementation of HUXt. The core codes of HUXt are contained in \emph{huxt.py}, and use of HUXt depends on three classes; \textbf{HUXt}, \textbf{ConeCME}, and \textbf{Observer}. Figure \ref{fig:huxt_diagram} presents a diagram of these classes and their methods. In this diagram, arrows indicate the passing of information from instances of one class to another.

Instantiating an instance of the \textbf{HUXt} class configures and initialises the model. There is a minimum required input of specifying the inner boundary condition with an array of solar wind speeds. Different examples of how these can be derived are presented below. Three main methods are attached to the \textbf{HUXt} class, \textbf{HUXt.solve}, \textbf{HUXt.save}, and \textbf{HUXt.get\_observer}. The \textbf{HUXt.solve} method sets the model running with optional inputs such as CMEs and streakline tracing, after which the results will be stored as \textbf{HUXt} attributes. In this way it is possible and practical to work with HUXt both programatically and interactively. The \textbf{HUXt.save} method writes all the data stored in attributes to an HDF5 file. The \textbf{HUXt.ts\_from\_vlong} method is used to convert a Carrington longitude profile of solar wind speeds into an time series of solar wind speeds at the HUXt inner boundary, under the assumption of synodic or sidereal rotation of the inner boundary. Whilst the \textbf{HUXt.get\_observer} links to the \textbf{Observer} class to provide ephemeris data on a range of solar system bodies interpolated onto the model time grid. 

The \textbf{Observer} class provides access to ephemeris data for Mercury, Venus, Earth, Mars, Jupiter, and Saturn as well as STEREO-A and STEREO-B, for the period spanning 1963-01-01 to 2029-01-01 at 4 hours resolution. The \textbf{Observer} class requires as input the name of the body and an array of times to output the ephemeris data. The ephemeris are linearly interpolated onto the output times from the 4 hour resolution data, and positions are provided in the Heliocentric Earth Equatorial (HEEQ), Heliocentric Aries Ecliptic (HAE), and Carrington coordinate systems.

Cone CMEs are represented by their own class, \textbf{ConeCME}. This class requires as input the 6 Cone CME parameters of source longitude, source latitude, full angular width, CME initial speed, CME radial thickness, and launch time relative to the model initiation time. Including Cone CMEs in a HUXt simulation is performed by passing a list of \textbf{ConeCME} objects to the \textbf{HUXt.solve} method. Attributes of the \textbf{ConeCME} class describe all of the CME's properties and its coordinates throughout the HUXt solution. There is also a method attached to the \textbf{ConeCME} class to compute the CME arrival time at any of the solar system bodies included in the HUXt ephemeris data. The \textbf{ConeCME.parameter\_array} method returns a numpy array of the Cone CME parameters, which is primarily intended to pass the CME parameters to the Numba optimised numerical core. This is required as the \textbf{HUXt} and \textbf{ConeCME} attributes relating to physical parameters are stored as Astropy quantities with the associated units, which are not interoperable with Numba optimised functions. More details on the updated implementation and use of Cone CMEs are discussed section \ref{sec:conecme}.

Basic support for 3-D simulations is provided as part of the \textbf{HUXt3d} class. In essence, this class is a wrapper around a collection of \textbf{HUXt} classes; one for each latitude simulated. Cone CMEs can be included using the same syntax as with the \textbf{HUXt} class. At present time-dependent boundary conditions are not fully supported for the \textbf{HUXt3d} container class, but all standard data and methods can be accessed for the individual \textbf{HUXt} inner classes in the standard manner. 

\subsection{Test suite}
In the latest release of HUXt, v4.0.0, a small test suite has been included in \emph{test\_huxt.py}. The aim of the test suite is to help ensure consistency and robustness of HUXt simulations for different users, as well as across future versions of HUXt. The test suite uses the \emph{pytest} testing framework and, at present, includes four tests. The first test is based on the comparison of the HUXt numerical solution and the analytical solution for the simple scenario of a constant inner boundary condition, as discussed in \ref{sec:testing} and Figure \ref{fig:analytical_error}. Therefore, this test helps ensure that the numerical scheme is working as expected. 

The second test compares a HUXt simulation of a Cone CME erupting into structured solar wind with a reference simulation of the same scenario, where these reference data are included as part of HUXt. The solar wind solution, as well as the Cone CME properties and arrival time at Earth are checked for consistency. This test helps to ensure that the core functionality of HUXt is producing consistent results for different users, on different systems, and across different future versions.

The third test similarly compares output with reference data, but to test the reproduction of streakline tracing, described in Section\ref{sec:streaklines}.

Finally, the fourth test compares solutions using an inner boundary at $30~R_{\odot}$  with the same boundary conditions backmapped to $10~R_{\odot}$, as described in Section \ref{sec:backmap}. While solutions are not expected to be identical, we define the acceptable tolerance for resulting conditions at 1 AU.

Whilst the current test suite goes some way to ensuring the core functionality of HUXt is performing reliably, at present not all of the HUXt functionality is supported by tests. Therefore it is a development priority to expand the test suite in future versions of HUXt.

\section{Functionality}
\label{sec:functionality}

\subsection{Updated functionality}
\subsubsection{Heliospheric extrapolations of coronal model output}

\begin{figure}[ht]
\begin{center}
\includegraphics[width=15cm]{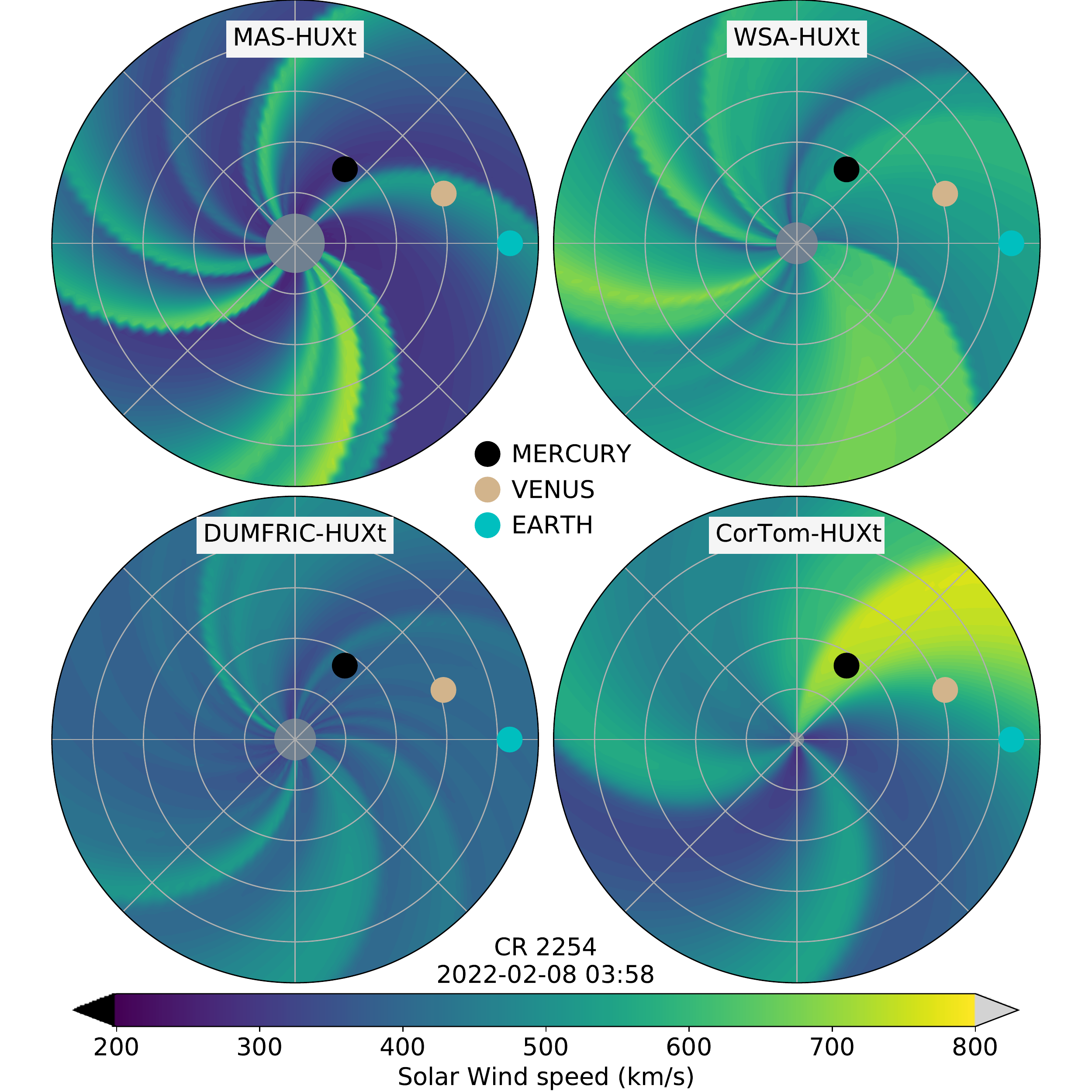}
\end{center}
\caption{HUXt solutions for Carrington rotation 2254, using inner boundary conditions taken from the MAS, WSA, DUMFRIC and CorTom models at the heliographic equator.}\label{fig:ambient_example}
\end{figure}

The primary function of HUXt is to provide a computationally efficient heliospheric extrapolation of the radial solar wind speed estimated by coronal models, such as, for example,  Wang-Sheeley-Arge (WSA) \cite{arge2000}, Magnetohydrodyanmics-About-A-Sphere (MAS) \cite{Riley2001} and the Durham Magnetofrictional (DUMFRIC) model \cite{yeates2010}. HUXt is agnostic concerning the input data series; it is not tuned or intended to be used with any particular source of boundary conditions. Figure \ref{fig:ambient_example} shows the HUXt solutions for Carrington rotation 2254 (beginning 2022-02-08) using boundary conditions taken from the MAS, WSA, and DUMFRIC coronal models, as well as boundary conditions derived from Coronal Tomography (CorTom)\cite{morgan2019,morgan2020}. MAS provides conditions at $30~R_{\odot}$, WSA and DUMFRIC produce conditions at $21.5~R_{\odot}$, while CorTom output is at $8~R_{\odot}$. HUXt is used with inner boundaries at these heights, without need to map the speeds to other radial distances. Whilst there is some agreement between the stream structure in these solutions, the absolute values and fine scale structure are significantly different. This serves to highlight the impact different model assumptions and architectures can have on the resulting estimate of the state of the corona, and consequently the state of the heliosphere. A detailed analysis of which factors determine the differences between these boundary conditions is outside the scope of this article. But we note that these models each make different approximations regarding the physics governing the structure of the coronal magnetic field; WSA approximates the coronal magnetic field as a potential magnetic field; DUMFRIC approximates the coronal magnetic field as a non-potential field; MAS approximates the coronal state using MHD. Furthermore, for WSA, DUMFRIC, and MHD, differences can arise due to the source and processing of the required magnetogram data \cite{gonzi2020}. CorTom is more fundamentally distinct from MAS, WSA and DUMFRIC, being based on a tomographic reconstruction of the coronal mass density derived from coronagraph data. \cite{gonzi2020} examined the sensitivity of ENLIL simulations to WSA and DUMFRIC inner boundary conditions. We think a future study that examines the differences between WSA, DUMFRIC, MAS and CorTom  derived boundary conditions and their impact on heliospheric simulations would be a valuable addition to the literature.

We also recognise the possibility of using HUXt -- which can investigate a large parameter space rapidly -- to calibrate the coronal model solar wind speeds \cite{mcgregor2011} in a manner which accounts for heliospheric acceleration and stream interactions. For example, \cite{bunting2022} used HUXt in the calibrating an empirical relationship that converts the CorTom coronal densities into solar wind speeds.

Regarding the ambient solar wind boundary conditions, several convenience functions are provided in \textit{huxt\_inputs.py} for loading and processing solar wind speed data from the MAS, WSA, PFSS and DUMFRIC models and CorTom outputs.  

\subsubsection{Cone CMEs}
\label{sec:conecme}
CMEs are incorporated in HUXt via the Cone CME parameterisation, wherein CMEs are treated as a velocity perturbation at the model inner boundary. Cone CMEs are purely hydrodyanmic structures, and have no magnetic field structure. In Cartesian space the shape of Cone CME is formed by two hemispheres connected by a cylindrical section, akin to a short sausage, with a limiting case of a sphere. A detailed description of the Cone CME geometry is presented in \cite{na2017}. This structure is directed radially away from the Sun and advects through the model inner boundary at the CME speed. Any location on the model inner boundary that intersects the Cone CME volume is assigned the CME speed. A Cone CME is fully specified by 6 parameters; longitude and latitude in HEEQ coordinates, full angular width, speed, thickness (radial length of the cylindrical section), and the launch time relative to the model initialisation time. Conversion of CME coordinates for use with synodic and sidereal frames is handled automatically. Functionality exists for importing a standard 'Cone CME' input file, such as produced by the UK Met Office CAT tool.

In HUXt v1.0, the position of a Cone CME was tracked by comparing simulations with and without the Cone CME and extracting the boundaries of regions where the simulations differed by more than $20 km~s^{-1}$. This approach was generally successful but could struggle to identify the boundary of Cone CMEs with speeds similar to the ambient solar wind speed. To improve upon this, we experimented with tracking the Cone CMEs using a discretised tracer field, but this was found to be too diffusive in practice and consequently required arbitrary thresholds to determine the CME boundaries. Therefore, from v2.0 onwards, Cone CMEs are tracked through the HUXt solution using individual tracer particles that follow the leading and trailing edge of the CME. These tracer particles are inserted into the flow onto the CME boundary as it advects through the model inner boundary and the tracer particles then passively advect through the flow solution. The tracer particles are injected onto every longitudinal and latitudinal coordinate that intersects the ConeCME. An outline of the tracer particle advection algorithm is given in Algorithm \ref{alg:advection}. The CME boundary is computed automatically at each time-step from the locations of the tracer particles. This method of tracking the CME boundary performs well for all CME speeds, and so is a significant improvement over the tracker function in v1.0.

\begin{algorithm}
	\caption{Tracer particle advection} 
    \label{alg:advection}
	\begin{algorithmic}[1]
		\For {each model time, with time step $\Delta t$}
            \State Update the model solution.
            \For {each test particle}
                \State Interpolate the flow speed at the test particle's current location, $V_{int}$
                \State Advance the test particle's radial coordinate by $V_{int}\Delta t$.
                \State Update array of time history of test particle coordinates.
            \EndFor
        \EndFor
	\end{algorithmic} 
\end{algorithm}

\begin{figure}[ht]
\begin{center}
\includegraphics[width=15cm]{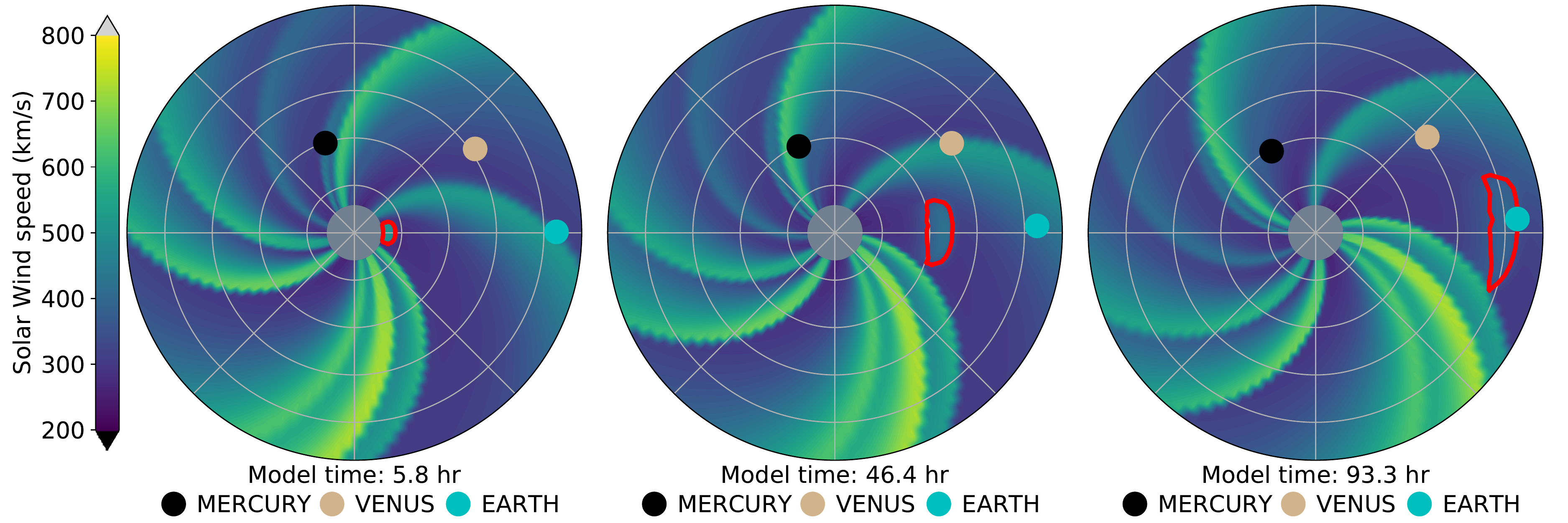}
\end{center}
\caption{Snapshots a HUXt solution including a cone CME. The ambient solar wind boundary condition was taken from MAS for Carrington rotation 2254, as in Figure \ref{fig:ambient_example}, while the Cone CME parameters represent a climatologically average CME. From left to right, the panels show the Cone CME (in red) shortly after initiation, halfway through its transit to Earth, an on arrival at Earth.
}\label{fig:cone_cme_example}
\end{figure}

Figure \ref{fig:cone_cme_example} shows snapshots from an example HUXt simulation including a Cone CME, with the red line marking the boundary of the CME. The Cone CME parameters were set to represent an Earth directed climatological average CME, with initial speed of $495 km~s^{-1}$ and full width of $37.4^{\circ}$, where these values were the median speed and width values from the KINCAT catalogue of CME parameters in the HELCATS database \cite{barnes2020}. Using the \textbf{ConeCME.compute\_arrival\_at\_body()} method, we calculated that the CME in the simulation in Figure \ref{fig:cone_cme_example} arrived at Earth after approximately $93.9$~hours, with an arrival speed of $360~km~s^{-1}$.

\subsubsection{Analysis, Figures, and Animations}
\label{sec:graphics}

A range of basic analysis and plotting capabilities are provided in \textit{huxt\_analysis.py}. These capabilities are demonstrated throughout the figures in this paper. Support is provided for:

\begin{itemize}
    \item Plotting time vs speed at fixed spatial coordinate.
    \item Plotting radius vs speed at fixed longitude and time. 
    \item Polar plot of speed as function of radius and longitude at fixed time and latitude.
    \item Polar plot of speed as function of radius and latitude at fixed time and longitude.
    \item Extracting time series of model parameters at bodies included in the \textbf{Observer} class.
    \item Comparison of HUXt simulations with NASA's OMNI data.
\end{itemize}

Furthermore,  both the latitudinal (e.g. Figure \ref{fig:ambient_example}) and longitudinal (right-hand panel of Figure \ref{fig:huxt3d}) cuts through the model can be animated over the model run duration using functions provided in \textit{huxt\_analysis.py}. Comparison of HUXt simulations with the NASA's OMNI data \cite{king2005} is facilitated via the on-demand download of OMNI data using SunPy's FIDO functionality \cite{barnes2020b}.

\subsection{New functionality}

\subsubsection{In-situ boundary conditions}
\label{sec:insitubc}
HUXt accepts fully time-dependent boundary conditions by specifying solar wind speed (and magnetic field polarity, see section \ref{sec:streaklines}) as a function of Carrington longitude and time at the inner boundary. In principle, this allows time-dependent coronal model output to be utilised, though this has yet to be fully tested. The second use case is initialisation of HUXt with in-situ solar wind measurements that are gridded into a Carrington longitude map. Functions for downloading the necessary OMNI data on-demand and generating the inner boundary condition, as well as setting up the HUXt model, are provided in \textbf{huxt\_inputs.py}. Simple corotation smoothed back and forward in time \cite{owens2021} is used to construct the boundary conditions. HUXt will accept inputs from more advanced methods, such as dynamic time warping \cite{owens2021}, though such processing is not included as part of the HUXt codebase and can instead be accessed at \url{https://github.com/University-of-Reading-Space-Science/SolarWindInputs_DTW}.

\begin{figure}[ht]
\begin{center}
\includegraphics[width=15cm]{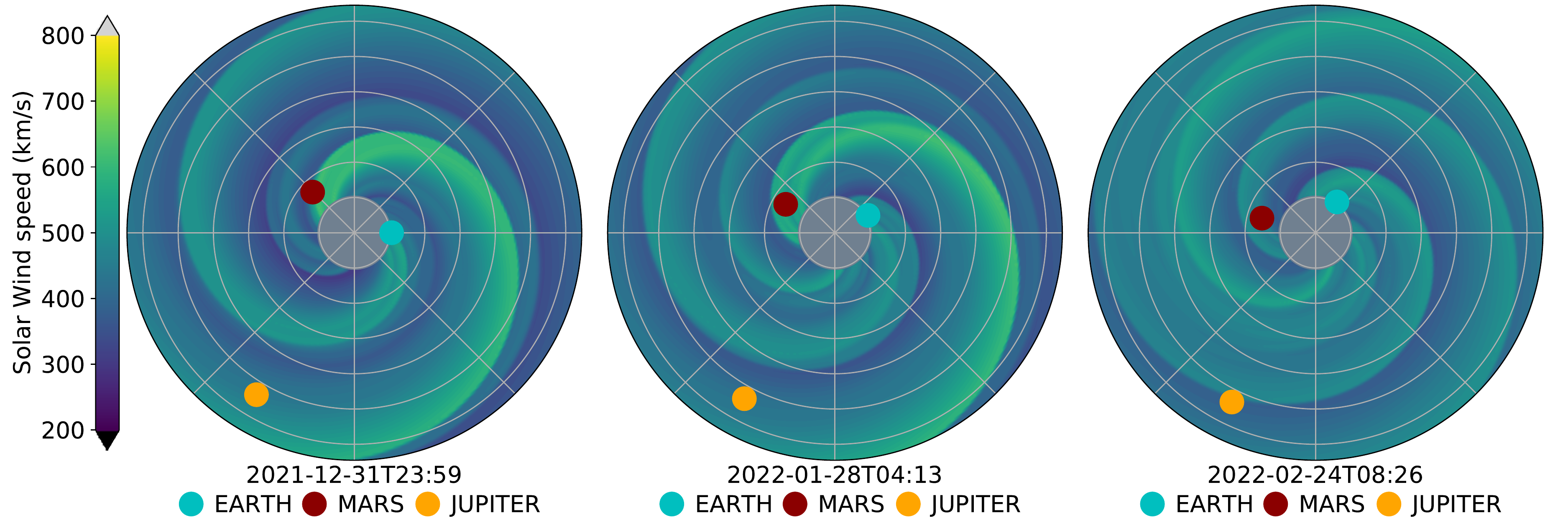}
\end{center}
\caption{(Snapshots from a HUXt simulation of a radial domain spanning 1~AU to 6~AU, initialised with the OMNI observations of in-situ solar wind conditions near-Earth. The snapshots are taken approximately 27 days apart.}\label{fig:omni_boundary_example}
\end{figure}

Figure \ref{fig:omni_boundary_example} shows an example of a HUXt simulation of the region from 1~AU to 6~AU, with the inner boundary condition derived from 4 months of near-Earth solar wind speed observations in the OMNI database \cite{king2005}. These three snapshots are each approximately one solar rotation (27 days) apart, meaning for time-independent boundary conditions, they would be identical. Here, the time evolution of the ambient solar wind structure can clearly be seen, with the fast solar wind streams decreasing in strength from late 2021 into early 2022. Cone CMEs can be included in such time-dependent boundary condition runs, in the same manner as for `standard' HUXt runs.

\subsubsection{Streakline tracing and sector structure mapping}
\label{sec:streaklines}
Functionality exists for tracing streaklines through HUXt flow fields. A streakline is the locus formed by connecting the locations of fluid parcels that originated or passed through a particular location, for example, the curve formed by smoke flowing from a chimney \cite{batchelor2000}. In HUXt, streaklines are computed by advecting test particles through the flow field from a fixed Carrington longitude. The algorithm for tracking a streakline from a fixed Carrington longitude is similar to tracking the Cone CME boundaries and hence depends primarily on Algorithm \ref{alg:advection}. The difference between the streakline tracing and the CME boundary tracing is in computing when and where the tracer particles are initialised. For the streaklines, this is done by computing the set of model time and longitude coordinates corresponding to when a specified Carrington longitude rotates into a model longitude bin. Because of the frozen-in-flux theorem, and under the assumption that the magnetic field is passive, the computed streakline will approximate a Parker spiral magnetic field line.

\begin{figure}[ht]
\begin{center}
\includegraphics[width=15cm]{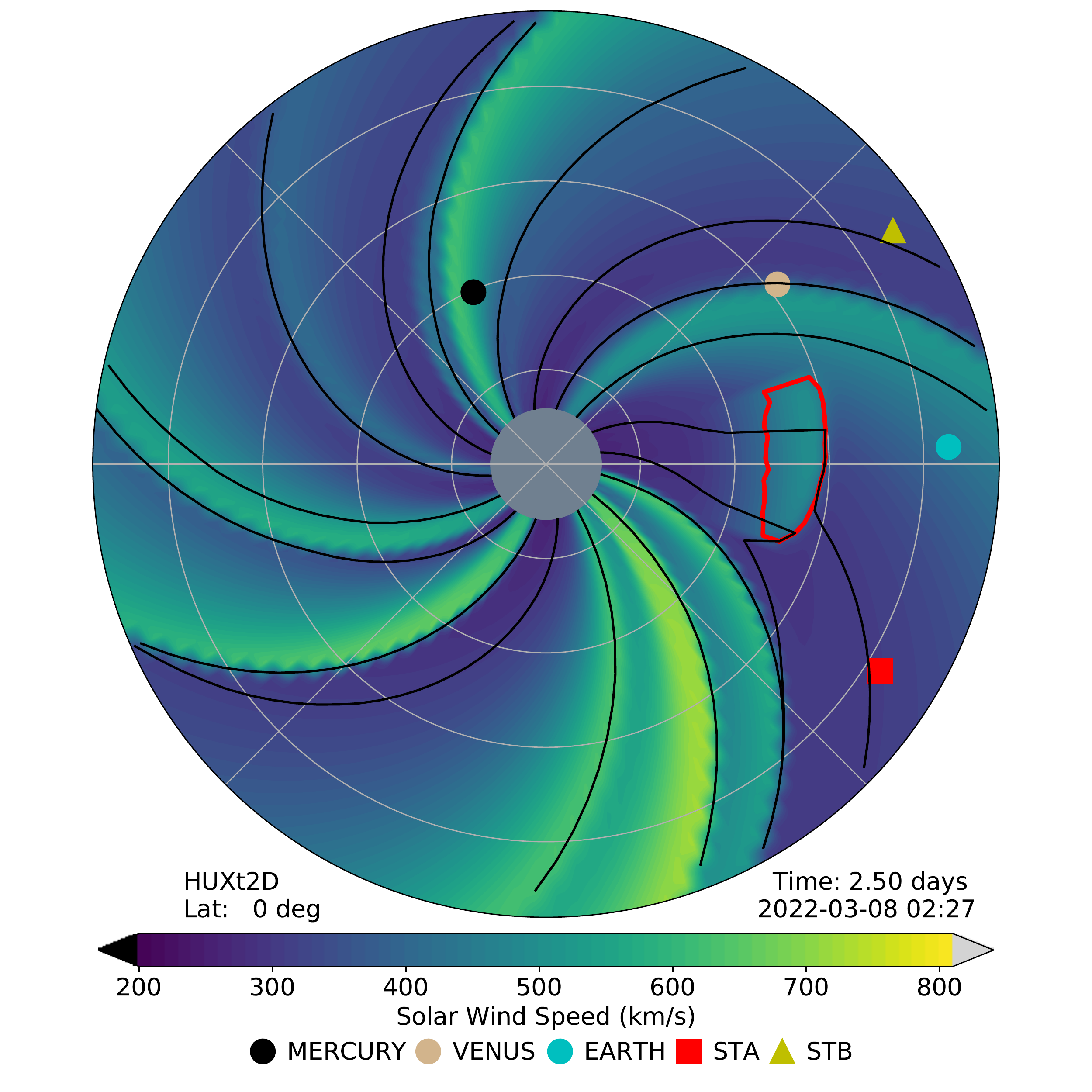}
\end{center}
\caption{A snapshot of the HUXt simulation of Carrington rotation 2254 with streaklines (black lines) plotted for every $22.5^{\circ}$ of Carrington longitude. The boundary of a cone CME is shown in red.}\label{fig:streakline_example}
\end{figure}

Figure \ref{fig:streakline_example} shows a snapshot from a HUXt simulation of CR2254 with streaklines plotted that originate from every $22.5^{\circ}$ of Carrington longitude. As expected, the streaklines follow the expected Parker spiral pattern, with `field lines' in faster flow regions being less tightly wound than in slower flow \cite{Owens2013a}. An Earth-directed cone CME has been inserted to show the disruption of the Parker spiral, with draping of the `field' across the CME front.

Previous versions of HUXt used the radial magnetic field polarity at the inner boundary to track magnetic sector structure as a discretised passive tracer field. Unfortunately, this approach proved too diffusive, with narrow sectors being eroded away, particularly for long-duration, outer heliosphere simulations.

\begin{figure}[ht]
\begin{center}
\includegraphics[width=15cm]{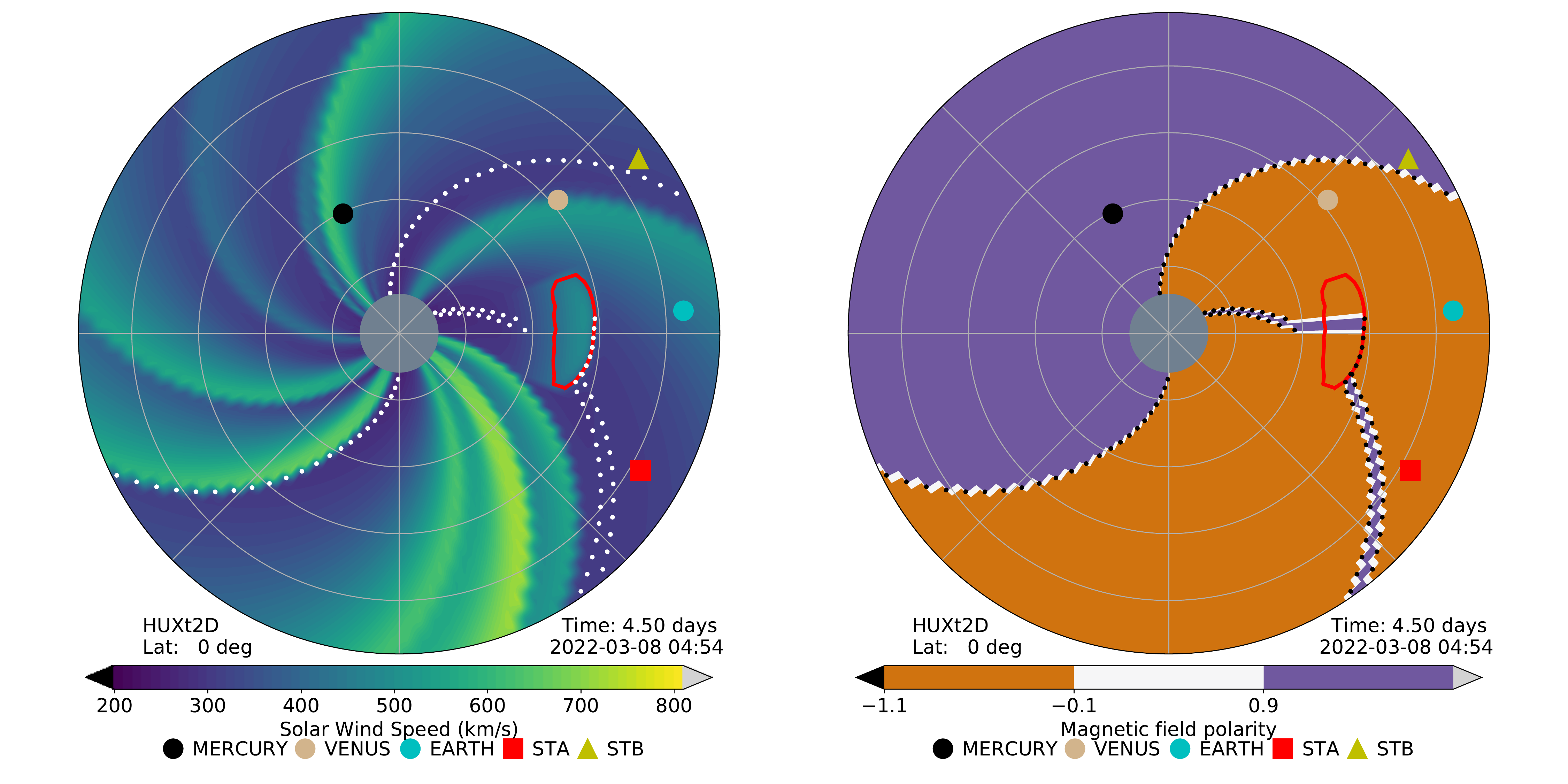}
\end{center}
\caption{Snapshots of the HUXt solar wind speed (left) and magnetic field polarity (right) for Carrington rotation 2254. The boundary of a cone CME is shown in red. Black and white lines show heliospheric current sheet positions of positive to negative radial field transitions (with increasing radial distance) and negative to positive, respectively.}\label{fig:polarity_example}
\end{figure}

The streakline functionality can be used to more effectively track the position of Carrington longitudes of interest through the solar wind, such as the heliospheric current sheet (HCS). Changes in polarity of the radial magnetic field from a coronal model, such as MAS or WSA, can be traced out through the HUXt flow field. This is shown in the left-hand panel of Figure \ref{fig:polarity_example}. The HCS locations which mark the transition from positive to negative radial field (with increasing radial distance) are shown by white lines, while the converse are shown by black lines. Then, the polarity map is found by associating regions of the model domain between the streaklines of the HCS with the appropriate polarity, shown in the right-hand panel of Figure \ref{fig:polarity_example}.  While this period, CR2254, shows a predominantly two-sector magnetic structure, there is a very short pair of HCS crossings around Earth longitude. Despite their proximity, these are preserved by the streakline method. It can also be seen how the addition of a cone CME disrupts the normal Parker spiral pattern of the HCS.

\subsubsection{Backmapping}
\label{sec:backmap}

Although the default inner boundary of HUXt is $30~r_{\odot}$, it is easily configured to use different inner boundary heights. This facilitates easier comparisons with a range of observables, as well as models initialised at other radial distances - for example, the commonly used $21.5~r_{\odot}$ inner boundary. In these circumstances, it can be necessary to map inner boundary conditions at one altitude to another. For example, mapping the input solar wind speed boundary condition from $30~r_{\odot}$ down to $15~r_{\odot}$. 

This is a non-trivial calculation, because it is necessary to account for the expected solar wind acceleration between the original and desired altitudes. HUXt provides a function to compute this mapping. For a solar wind parcel on the initial boundary height, this function computes both the expected speed and source longitude of this parcel at the desired altitude, in a manner consistent with the HUXt model dynamics. The derivation of the equations used to compute this mapping are provided in Appendix \ref{apndx:backmap}. Note that stream interactions are ignored for backmapping, though this effect is expected to be very small close to the Sun.

\begin{figure}[ht]
\begin{center}
\includegraphics[width=15cm]{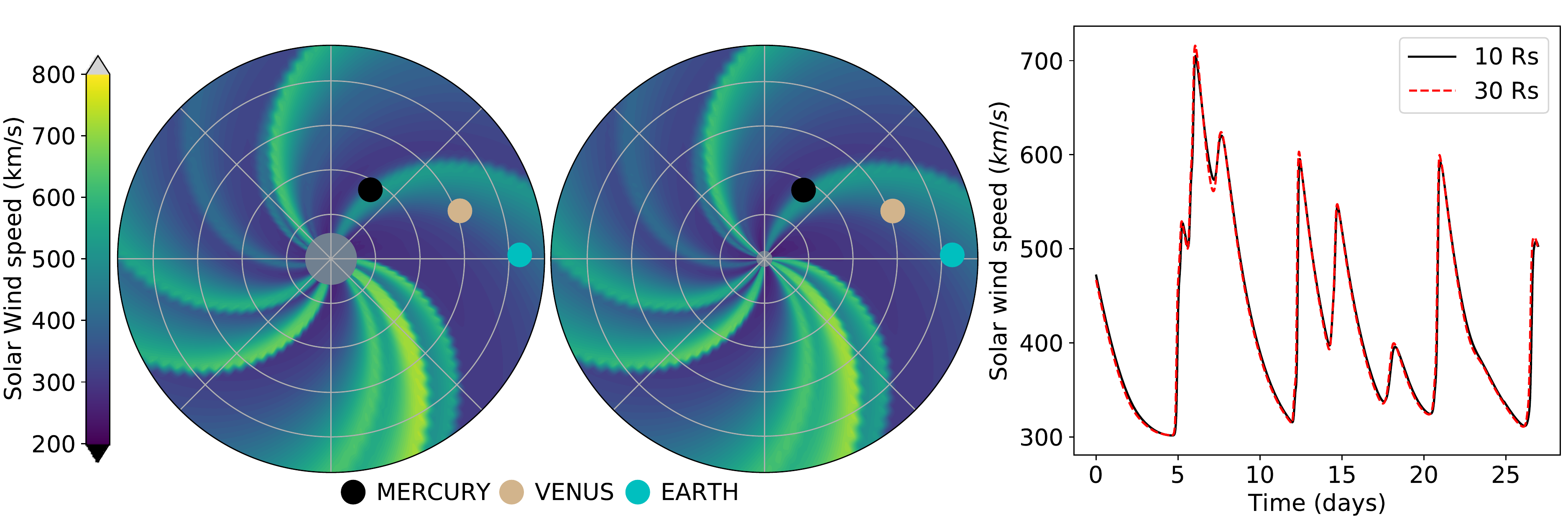}
\end{center}
\caption{(Left) A snapshot of the HUXt simulation of Carrington rotation 2254 initialised at $30~R_{\odot}$. (Middle) A snapshot of a HUXt simulation of Carrington rotation 2254 where the inner boundary condition has been backmapped to $10~R_{\odot}$. (Right) Time series of the solar wind speed at Earth from the HUXt simulations initialised at $30~R_{\odot}$ (red line) and $10~R_{\odot}$ (black line)}\label{fig:backmap_example}
\end{figure}

Figure \ref{fig:backmap_example} shows an example of the results of the backmapping procedure. The left panel shows a HUXt simulation initialised at $30~R_{\odot}$ with data from a MAS simulation of CR2254. The middle panel shows a HUXt simulation where the inner boundary was backmapped to $10~R_{\odot}$. The right panel shows the resulting solar wind speed time series at Earth for 27 days. The time series are very similar, but the gradients and magnitudes are slightly smaller in the solution initialised at $10~R_{\odot}$. This is an artefact of the backmapping procedure, which cannot distinguish between solar wind parcels of different speeds that have distinct source longitudes at a larger radii, but are estimated to have the same source longitude at a smaller radii. These overlapping parcels must necessarily be averaged and interpolated onto the regular HUXt longitude grid, which serves to smooth the inner boundary condition. It is important to be aware of this impact when backmapping an inner boundary condition to a different height.

Note that Cone CME parameters are defined relative to the model inner boundary. Thus moving the inner boundary closer to the Sun will require an earlier Cone CME insertion time and will likely need an increased initial speed, to counteract the addition deceleration of the increased propagation path. As the duration of the Cone CME speed perturbation at the inner boundary is set by the radius of the spherical disturbance, it may also be necessary to increase the Cone CME thickness in order to obtain the same speed perturbation far from the Sun.

\subsubsection{3-D solutions}
\label{sec:3d}

\begin{figure}[ht]
\begin{center}
\includegraphics[width=15cm]{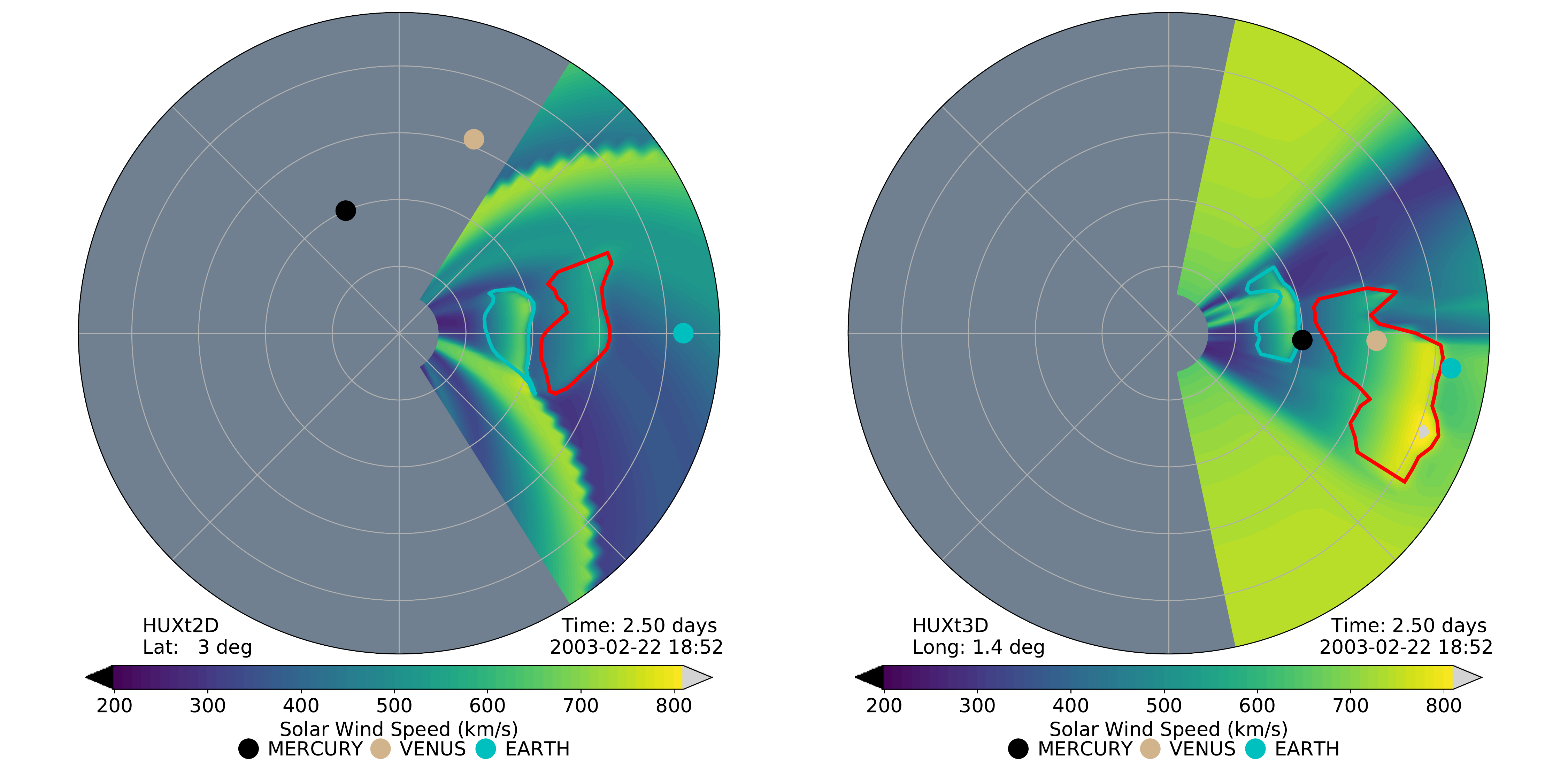}
\end{center}
\caption{A snapshot of a HUXt3D simulation from February 2003. (left) A radius-longitude cut at Earth latitude  and (right) a radius-latitude cut at Earth longitude. The boundaries of two cone CMEs are shown in red and cyan.}\label{fig:huxt3d}
\end{figure}

Figure \ref{fig:huxt3d} shows snapshots from a \textbf{HUXt3D} simulation of Carrington rotation 2000, from February 2003. The right-hand panel of Figure \ref{fig:huxt3d} shows a snapshot of the radial-latitudinal plane containing Earth. The left-hand panel of Figure \ref{fig:huxt3d} shows the ecliptic plane (i.e. the radial-longitude plane at a constant latitude closest to Earth's latitude). This simulation shows a somewhat typical declining/minimum solar cycle phase structure of fast wind at the poles and slower wind at mid-to-low latitudes. Two cone CMEs have been inserted, with their boundaries shown by the cyan and red lines.

\section{Summary and Future work}
\label{sec:Summary}

HUXt is an open source reduced physics numerical model of the solar wind, built in Python. The primary purpose of HUXt is to provide a solar wind model that is very computationally cheap and is of minimal complexity, so that it can serve as a surrogate for 3d-MHD solar wind models in context where such simulations are currently impractical, for example, large ensemble simulations and data asssimilative applications.

Version 4.0 includes two key improvements to the models functionality. Firstly, HUXt now accepts fully time-dependant boundary conditions, allowing HUXt to be run by time-series of solar wind speed observations from in-situ monitors, such as those provided by the OMNI dataset. Secondly, methods have been included to compute flow streaklines of the HUXt simulations. With these streaklines it is then possible to estimate the location of the Heliospheric Current Sheet, and produce maps radial heliospheric magnetic field polarity.

Additionally, this version is the first to incorporate a small test-suite, which tests the HUXt numerical scheme against a simple analytical solution, and checks a HUXt simulation for consistency with some reference simulation data. This is a first step to improving the reliability and reproducibility of HUXt. Expanding this test suite to cover all of HUXt's functionality is a development priority for future versions.

\subsection{Data Assimilation}

Data assimilation (DA) is the process of combining observations and models, accounting for the uncertainty in both, to achieve an optimal estimate of the state of a system \cite{vanleeuwen2009}. It is widely used in meteorology to dramatically improve forecasting skill \cite{kalnay2005}, as well as in a number of areas of space physics \cite{brun2007,hickmann2015,Murray2015, chartier2016, Lang2017}. There are two broad categories of DA methods, variational methods and sequential methods. Variational methodologies, such as 4DVar, aim to find the most probable system state by processing all the observations simultaneously; whereas sequential methodologies, such as the particle filters, aim to find the minimal-variance state by processing observations one-by-one, in a sequential fashion \cite{Lang2017}. HUXt is of value to solar wind DA for two reasons.

Firstly, as it is computationally cheap, HUXt is amenable to ensemble DA methods, which require running 10s to 1000s of instances of a model for each analysis, be this for research purposes or in a forecasting context \cite{Lang2017}. For example, sequential DA methods, such as particle filters \cite{vanleeuwen2009}, use an ensemble of simulations to return an estimate of the state of a dynamical system. DA systems such as this are simple to implement and applicable in a broader range of contexts than some other DA methods (e.g. non-linear systems), but require an ensemble of particles that increases with the dimension of the state space of the model. Consequently, this can become very expensive to compute and becomes quickly impractical for models with large domains and/or numbers of parameters. HUXt is well suited to use with these methods, and we are developing a particle filter DA scheme for assimilating time-elongation profiles of CMEs observed by white-light imagers such as coronagraphs and heliospheric imagers.

Variational DA methods, such as 4DVar, often require the formulation of an 'adjoint' model, with which a models sensitivities to perturbations can be assessed. The relatively simplicity of the HUXt equations and code base means that the `adjoint' model can be constructed with (relative) ease. This enables powerful variational DA methods to be employed \cite{Lang2019}, which are proving very promising for assimilating in situ solar wind observations \cite{lang2021}. 

\subsection{SWIMMR}

The UK government is funding the improvement of its national space weather forecasting capability, particularly through the targeted Space Weather Instrumentation, Measurement, Modelling and Risk (SWIMMR) programme. Within SWIMMR, HUXt is being developed for operational use at the UK Met Office Space Weather Operations Center, as part of the Space Weather Empirical Ensembles Package (SWEEP). The motivation for SWEEP is that the dominant source of uncertainty in numerical model predictions and forecasts of heliosperhic solar wind speed are the uncertainties at the inner boundary of these models.

SWEEP aims to better quantify these uncertainties by producing a large ensemble of forecasts driven with boundary conditions derived from independent data sources (e.g. magnetograms and white light coronal observations) and different physical assumptions, e.g. potential and non-potential coronal magnetic fields. In doing so, SWEEP will produce ensemble forecasts with improved real-world representivity. This work depends critically on having a computationally cheap numerical model like HUXt, as such large ensembles across multiple inputs can not yet be practically generated with operationally used 3D MHD models. SWEEP is expected to be operational by 2024.

\section{Appendix A: Backmapping equations}
\label{apndx:backmap}
Substituting equation \ref{eq:acc_term} into \ref{eq:vsw_eqn}, we obtain

\begin{equation}
V = V_0 + \alpha V_0 - \alpha V_{0}e^{\left(\frac{r_0 - r}{r_h} \right)}
\end{equation}

The travel time for a solar wind parcel to propagate between and inner radius, $r_{in}$, to an outer radius, $r_{out}$, is 

\begin{equation}
T = \int _{r_{in}}^{r_{out}} \frac{1}{V}dr
\end{equation}

Defining $A = V_{0} + \alpha V_{0}$, this integral becomes

\begin{equation}
T = \int _{r_{in}}^{r_{out}} \frac{1}{A-\alpha V_{0}e^{\left(\frac{r_0 - r}{r_h}\right)}}dr
\end{equation}

for which the solution is 

\begin{equation}
T = \left[\frac{r_h log\left(Ae^{\left(\frac{r}{r_h}\right)} - \alpha V_{0}e^{\left(\frac{r_0}{r_h}\right)}  \right)}{A} \right]_{r_{in}}^{r_{out}}
\end{equation}

This travel time can be used in conjunction with the synodic solar rotation rate to estimate the source longitude of a solar wind parcel at an inner radius relative to its source longitude at the outer radius.

\section*{Conflict of Interest Statement}
The authors declare that the research was conducted in the absence of any commercial or financial relationships that could be construed as a potential conflict of interest.

\section*{Author Contributions}
LB and MO contributed to the conception of this article. MO conceived the HUXt model, whilst both MO and LB co-lead the development of HUXt. Both authors contributed to the writing and revision of the submitted article.

\section*{Funding}
This work was part-funded by Science and Technology Facilities Council (STFC) grant numbers ST/R000921/1 and ST/V000497/1, and NERC grant number NE/S010033/1.

\section*{Acknowledgements}
This research made use of Astropy (\url{http://www.astropy.org}), a community-developed core Python package for Astronomy \cite{robitaille2013, price-whelan2018}.

This research used version 4.0.0 \cite{mumford2022} of the SunPy open source software package \cite{barnes2020b}.

Figures for this article were made with version 3.3.4 of Matplotlib \cite{caswell2021,hunter2007}

We have benefited greatly from the availability of the large archive of Predictive Science Inc's MAS coronal model solutions (\url{https://www.predsci.com/mhdweb/home.php}) for much of the HUXt testing and development and thank Pete Riley for useful discussions on solar wind modelling.

\section*{Code and Data Availability Statement}
HUXt is available through both Zenodo \cite{barnard2021} and GitHub (\url{https://github.com/University-of-Reading-Space-Science/HUXt}). The code that generated the results and figures in this article are available as Jupyter notebooks hosted at  \url{https://github.com/University-of-Reading-Space-Science/HUXt_Frontiers_article}.


\begin{thebibliography}{50}
\providecommand{\natexlab}[1]{#1}
\expandafter\ifx\csname urlstyle\endcsname\relax
  \providecommand{\doi}[1]{doi:\discretionary{}{}{}#1}\else
  \providecommand{\doi}{doi:\discretionary{}{}{}\begingroup
  \urlstyle{rm}\Url}\fi
\providecommand{\selectlanguage}[1]{\relax}
\providecommand{\bibAnnoteFile}[1]{%
  \IfFileExists{#1}{\begin{quotation}\noindent\textsc{Key:} #1\\
  \textsc{Annotation:}\ \input{#1}\end{quotation}}{}}
\providecommand{\bibAnnote}[2]{%
  \begin{quotation}\noindent\textsc{Key:} #1\\
  \textsc{Annotation:}\ #2\end{quotation}}

\bibitem[{Arge and Pizzo(2000)}]{arge2000}
Arge, C.~N. and Pizzo, V.~J. (2000).
\newblock Improvement in the prediction of solar wind conditions using
  near-real time solar magnetic field updates.
\newblock \emph{Journal of Geophysical Research: Space Physics} 105,
  10465--10479.
\newblock \doi{10.1029/1999JA000262}
\input{arge2000}

\bibitem[{Barnard and Owens(2021)}]{barnard2021}
[Dataset] Barnard, L. and Owens, M. (2021).
\newblock University-of-{{Reading-Space-Science}}/{{HUXt}}: {{HUXt}}.
\newblock Zenodo.
\newblock \doi{10.5281/zenodo.4889327}
\input{barnard2021}

\bibitem[{Barnard et~al.(2020)}]{barnard2020}
Barnard, L., Owens, M.~J., Scott, C.~J., and de~Koning, C.~A. (2020).
\newblock Ensemble {{CME Modeling Constrained}} by {{Heliospheric Imager
  Observations}}.
\newblock \emph{AGU Advances} 1, e2020AV000214.
\newblock \doi{10.1029/2020AV000214}
\input{barnard2020}

\bibitem[{Barnes et~al.(2020{\natexlab{a}})}]{barnes2020}
Barnes, D., Davies, J.~A., Harrison, R.~A., Byrne, J.~P., Perry, C.~H.,
  Bothmer, V., et~al. (2020{\natexlab{a}}).
\newblock {{CMEs}} in the {{Heliosphere}}: {{III}}. {{A Statistical Analysis}}
  of the {{Kinematic Properties Derived}} from {{Stereoscopic Geometrical
  Modelling Techniques Applied}} to {{CMEs Detected}} in the {{Heliosphere}}
  from 2008 to 2014 by {{STEREO}}/{{HI-1}}.
\newblock \emph{Solar Physics} 295.
\newblock \doi{10.1007/s11207-020-01717-w}
\input{barnes2020}

\bibitem[{Barnes et~al.(2020{\natexlab{b}})}]{barnes2020b}
Barnes, W.~T., Bobra, M.~G., Christe, S.~D., Freij, N., Hayes, L.~A., Ireland,
  J., et~al. (2020{\natexlab{b}}).
\newblock The {{SunPy Project}}: {{Open Source Development}} and {{Status}} of
  the {{Version}} 1.0 {{Core Package}}.
\newblock \emph{The Astrophysical Journal} 890, 68.
\newblock \doi{10.3847/1538-4357/ab4f7a}
\input{barnes2020b}

\bibitem[{Batchelor(2000)}]{batchelor2000}
Batchelor, G.~K. (2000).
\newblock \emph{An {{Introduction}} to {{Fluid Dynamics}}}.
\newblock Cambridge {{Mathematical Library}} ({Cambridge}: {Cambridge
  University Press}).
\newblock \doi{10.1017/CBO9780511800955}
\input{batchelor2000}

\bibitem[{Brun(2007)}]{brun2007}
Brun, A.~S. (2007).
\newblock Towards using modern data assimilation and weather forecasting
  methods in solar physics.
\newblock \emph{Astronomische Nachrichten} 328, 329--338.
\newblock \doi{10.1002/asna.200610739}
\input{brun2007}

\bibitem[{Bunting and Morgan(2022)}]{bunting2022}
Bunting, K.~A. and Morgan, H. (2022).
\newblock An inner boundary condition for solar wind models based on coronal
  density.
\newblock \emph{Journal of Space Weather and Space Climate} 12, 30.
\newblock \doi{10.1051/swsc/2022026}
\input{bunting2022}

\bibitem[{Cannon et~al.(2013)}]{Cannon2013}
Cannon, P., Angling, M., Barclay, L., Curry, C., Dyer, C., Edwards, R., et~al.
  (2013).
\newblock \emph{Extreme Space Weather : Impacts on Engineered Systems and
  Infrastructure}.
\newblock Tech. rep., {Royal Academy of Engineering}
\input{Cannon2013}

\bibitem[{Case et~al.(2008)}]{case2008}
Case, A.~W., Spence, H.~E., Owens, M.~J., Riley, P., and Odstrcil, D. (2008).
\newblock Ambient solar wind's effect on {{ICME}} transit times.
\newblock \emph{Geophysical Research Letters} 35.
\newblock \doi{10.1029/2008GL034493}
\input{case2008}

\bibitem[{Caswell et~al.(2021)}]{caswell2021}
[Dataset] Caswell, T.~A., Droettboom, M., Lee, A., de~Andrade, E.~S., Hunter,
  J., Firing, E., et~al. (2021).
\newblock Matplotlib/matplotlib: {{REL}}: V3.3.4.
\newblock Zenodo.
\newblock \doi{10.5281/zenodo.4475376}
\input{caswell2021}

\bibitem[{Chartier et~al.(2016)}]{chartier2016}
Chartier, A.~T., Matsuo, T., Anderson, J.~L., Collins, N., Hoar, T.~J., Lu, G.,
  et~al. (2016).
\newblock Ionospheric data assimilation and forecasting during storms.
\newblock \emph{Journal of Geophysical Research: Space Physics} 121, 764--778.
\newblock \doi{10.1002/2014JA020799}
\input{chartier2016}

\bibitem[{Chi et~al.(2021)}]{chi2021}
Chi, Y., Scott, C., Shen, C., Barnard, L., Owens, M., Xu, M., et~al. (2021).
\newblock Modeling the {{Observed Distortion}} of {{Multiple}} ({{Ghost}})
  {{CME Fronts}} in {{STEREO Heliospheric Imagers}}.
\newblock \emph{The Astrophysical Journal Letters} 917, L16.
\newblock \doi{10.3847/2041-8213/ac1203}
\input{chi2021}

\bibitem[{Gonzi et~al.(2020)}]{gonzi2020}
Gonzi, S., Weinzierl, M., Bocquet, F.-X., Bisi, M.~M., Odstrcil, D., Jackson,
  B.~V., et~al. (2020).
\newblock Impact of {{Inner Heliospheric Boundary Conditions}} on {{Solar Wind
  Predictions}} at {{Earth}}.
\newblock \emph{Space Weather} , 1--40\doi{10.1029/2020SW002499}
\input{gonzi2020}

\bibitem[{Gopalswamy et~al.(2009)}]{gopalswamy2009}
Gopalswamy, N., Yashiro, S., Michalek, G., Stenborg, G., Vourlidas, A.,
  Freeland, S., et~al. (2009).
\newblock The {{SOHO}}/{{LASCO CME Catalog}}.
\newblock \emph{Earth, Moon, and Planets} 104, 295--313.
\newblock \doi{10.1007/s11038-008-9282-7}
\input{gopalswamy2009}

\bibitem[{Gosling(1993)}]{gosling1993}
Gosling, J.~T. (1993).
\newblock The solar flare myth.
\newblock \emph{Journal of Geophysical Research: Space Physics} 98,
  18937--18949.
\newblock \doi{10.1029/93JA01896}
\input{gosling1993}

\bibitem[{Hickmann et~al.(2015)Hickmann, Godinez, Henney, and
  Arge}]{hickmann2015}
Hickmann, K.~S., Godinez, H.~C., Henney, C.~J., and Arge, C.~N. (2015).
\newblock Data {{Assimilation}} in the {{ADAPT Photospheric Flux Transport
  Model}}.
\newblock \emph{Solar Physics} 290, 1105--1118.
\newblock \doi{10.1007/s11207-015-0666-3}
\input{hickmann2015}

\bibitem[{Hinterreiter et~al.(2021)}]{hinterreiter2021a}
Hinterreiter, J., Amerstorfer, T., Temmer, M., Reiss, M.~A., Weiss, A.~J.,
  M{\"o}stl, C., et~al. (2021).
\newblock Drag-{{Based CME Modeling With Heliospheric Images Incorporating
  Frontal Deformation}}: {{ELEvoHI}} 2.0.
\newblock \emph{Space Weather} 19, e2021SW002836.
\newblock \doi{10.1029/2021SW002836}
\input{hinterreiter2021a}

\bibitem[{Hunter(2007)}]{hunter2007}
Hunter, J.~D. (2007).
\newblock Matplotlib: {{A 2D Graphics Environment}}.
\newblock \emph{Computing in Science \& Engineering} 9, 90--95.
\newblock \doi{10.1109/MCSE.2007.55}
\input{hunter2007}

\bibitem[{Kalnay(2005)}]{kalnay2005}
Kalnay, E. (2005).
\newblock \emph{Atmospheric {{Modeling}}, {{Data Assimilation}} and
  {{Predictability}} | {{Atmospheric}} Science and Meteorology} ({Cambridge
  University Press}).
\newblock \doi{10.1017/CBO9780511802270}
\input{kalnay2005}

\bibitem[{King and Papitashvili(2005)}]{king2005}
King, J.~H. and Papitashvili, N.~E. (2005).
\newblock Solar wind spatial scales in and comparisons of hourly {{Wind}} and
  {{ACE}} plasma and magnetic field data.
\newblock \emph{Journal of Geophysical Research: Space Physics} 110.
\newblock \doi{10.1029/2004JA010649}
\input{king2005}

\bibitem[{Lang et~al.(2017)}]{Lang2017}
Lang, M., Browne, P., van Leeuwen, P.~J., and Owens, M.~J. (2017).
\newblock Data {{Assimilation}} in the {{Solar Wind}}: {{Challenges}} and
  {{First Results}}.
\newblock \emph{Space Weather} \doi{10.1002/2017SW001681}
\input{Lang2017}

\bibitem[{Lang and Owens(2019)}]{Lang2019}
Lang, M. and Owens, M.~J. (2019).
\newblock A {{Variational Approach}} to {{Data Assimilation}} in the {{Solar
  Wind}}.
\newblock \emph{Space Weather} 17, 59--83.
\newblock \doi{10.1029/2018SW001857}
\input{Lang2019}

\bibitem[{Lang et~al.(2021}]{lang2021}
Lang, M., Witherington, J., Turner, H., Owens, M.~J., and Riley, P. (2021).
\newblock Improving {{Solar Wind Forecasting Using Data Assimilation}}.
\newblock \emph{Space Weather} 19, e2020SW002698.
\newblock \doi{10.1029/2020SW002698}
\input{lang2021}

\bibitem[{Macneil et~al.(2022)}]{macneil2022}
Macneil, A.~R., Owens, M.~J., Finley, A.~J., and Matt, S.~P. (2022).
\newblock A statistical evaluation of ballistic backmapping for the slow solar
  wind: The interplay of solar wind acceleration and corotation.
\newblock \emph{Monthly Notices of the Royal Astronomical Society} 509,
  2390--2403.
\newblock \doi{10.1093/mnras/stab2965}
\input{macneil2022}

\bibitem[{Mays et~al.(2015)}]{Mays2015}
Mays, M.~L., Taktakishvili, A., Pulkkinen, A., MacNeice, P.~J., Rast{\"a}tter,
  L., Odstrcil, D., et~al. (2015).
\newblock Ensemble {{Modeling}} of {{CMEs Using}} the
  {{WSA}}\textendash{{ENLIL}}+{{Cone Model}}.
\newblock \emph{Solar Physics} 290, 1775--1814.
\newblock \doi{10.1007/s11207-015-0692-1}
\input{Mays2015}

\bibitem[{McGregor et~al.(2011)}]{mcgregor2011}
McGregor, S.~L., Hughes, W.~J., Arge, C.~N., Owens, M.~J., and Odstrcil, D.
  (2011).
\newblock The distribution of solar wind speeds during solar minimum:
  {{Calibration}} for numerical solar wind modeling constraints on the source
  of the slow solar wind.
\newblock \emph{Journal of Geophysical Research: Space Physics} 116.
\newblock \doi{10.1029/2010JA015881}
\input{mcgregor2011}

\bibitem[{Merkin et~al.(2016)}]{merkin2016}
Merkin, V.~G., Lyon, J.~G., Lario, D., Arge, C.~N., and Henney, C.~J. (2016).
\newblock Time-dependent magnetohydrodynamic simulations of the inner
  heliosphere.
\newblock \emph{Journal of Geophysical Research: Space Physics} 121,
  2866--2890.
\newblock \doi{10.1002/2015JA022200}
\input{merkin2016}

\bibitem[{Morgan(2019)}]{morgan2019}
Morgan, H. (2019).
\newblock An {{Atlas}} of {{Coronal Electron Density}} at
  {{5R}}\$\textbackslash less\$sub\$\textbackslash greater\$\$\o
  dot\$\$\textbackslash less\$/sub\$\textbackslash greater\$. {{II}}. {{A
  Spherical Harmonic Method}} for {{Density Reconstruction}}.
\newblock \emph{The Astrophysical Journal Supplement Series} 242, 3.
\newblock \doi{10.3847/1538-4365/ab125d}
\input{morgan2019}

\bibitem[{Morgan and Cook(2020)}]{morgan2020}
Morgan, H. and Cook, A.~C. (2020).
\newblock The {{Width}}, {{Density}}, and {{Outflow}} of {{Solar Coronal
  Streamers}}.
\newblock \emph{The Astrophysical Journal} 893, 57.
\newblock \doi{10.3847/1538-4357/ab7e32}
\input{morgan2020}

\bibitem[{Mumford et~al.(2022)}]{mumford2022}
[Dataset] Mumford, S.~J., Freij, N., Stansby, D., Christe, S., Ireland, J.,
  Mayer, F., et~al. (2022).
\newblock {{SunPy}}.
\newblock Zenodo.
\newblock \doi{10.5281/zenodo.6524764}
\input{mumford2022}

\bibitem[{Murray et~al.(2015)}]{Murray2015}
Murray, S.~A., Henley, E.~M., Jackson, D.~R., and Bruinsma, S.~L. (2015).
\newblock Assessing the performance of thermospheric modeling with data
  assimilation throughout solar cycles 23 and 24.
\newblock \emph{Space Weather} , n/a--n/a\doi{10.1002/2015SW001163}
\input{Murray2015}

\bibitem[{Na et~al.(2017)}]{na2017}
Na, H., Moon, Y.-J., and Lee, H. (2017).
\newblock Development of a {{Full Ice-cream Cone Model}} for {{Halo Coronal
  Mass Ejections}}.
\newblock \emph{The Astrophysical Journal} 839, 82.
\newblock \doi{10.3847/1538-4357/aa697c}
\input{na2017}

\bibitem[{Narechania et~al.(2021)}]{narechania2021}
Narechania, N.~M., Nikoli{\'c}, L., Freret, L., Sterck, H.~D., and Groth, C.
  P.~T. (2021).
\newblock An integrated data-driven solar wind \textendash{} {{CME}} numerical
  framework for space weather forecasting.
\newblock \emph{Journal of Space Weather and Space Climate} 11, 8.
\newblock \doi{10.1051/swsc/2020068}
\input{narechania2021}

\bibitem[{Odstrcil(2003)}]{Odstrcil2003}
Odstrcil, D. (2003).
\newblock Modeling 3-{{D}} solar wind structure.
\newblock \emph{Advances in Space Research} 32, 497--506.
\newblock \doi{10.1016/S0273-1177(03)00332-6}
\input{Odstrcil2003}

\bibitem[{Odstrcil et~al.(2020)}]{odstrcil2020}
Odstrcil, D., Mays, M.~L., Hess, P., Jones, S.~I., Henney, C.~J., and Arge,
  C.~N. (2020).
\newblock Operational {{Modeling}} of {{Heliospheric Space Weather}} for the
  {{Parker Solar Probe}}.
\newblock \emph{The Astrophysical Journal Supplement Series} 246, 73.
\newblock \doi{10.3847/1538-4365/ab77cb}
\input{odstrcil2020}

\bibitem[{Owens and Forsyth(2013)}]{Owens2013a}
Owens, M.~J. and Forsyth, R.~J. (2013).
\newblock The {{Heliospheric Magnetic Field}}.
\newblock \emph{Living Reviews in Solar Physics} 10.
\newblock \doi{10.12942/lrsp-2013-5}
\input{Owens2013a}

\bibitem[{Owens et~al.(2020)}]{owens2020a}
Owens, M.~J., Lang, M., Barnard, L., Riley, P., {Ben-Nun}, M., Scott, C.~J.,
  et~al. (2020).
\newblock A {{Computationally Efficient}}, {{Time-Dependent Model}} of the
  {{Solar Wind}} for {{Use}} as a {{Surrogate}} to {{Three-Dimensional
  Numerical Magnetohydrodynamic Simulations}}.
\newblock \emph{Solar Physics} 295, 43.
\newblock \doi{10.1007/s11207-020-01605-3}
\input{owens2020a}

\bibitem[{Owens and Nichols(2021)}]{owens2021}
Owens, M.~J. and Nichols, J.~D. (2021).
\newblock Using in situ solar-wind observations to generate inner-boundary
  conditions to outer-heliosphere simulations \textendash{} {{I}}. {{Dynamic}}
  time warping applied to synthetic observations.
\newblock \emph{Monthly Notices of the Royal Astronomical Society} 508,
  2575--2582.
\newblock \doi{10.1093/mnras/stab2512}
\input{owens2021}

\bibitem[{Owens and Riley(2017)}]{owens2017c}
Owens, M.~J. and Riley, P. (2017).
\newblock Probabilistic {{Solar Wind Forecasting Using Large Ensembles}} of
  {{Near-Sun Conditions With}} a {{Simple One-Dimensional}} ``{{Upwind}}''
  {{Scheme}}.
\newblock \emph{Space Weather} 15, 1461--1474.
\newblock \doi{10.1002/2017SW001679}
\input{owens2017c}

\bibitem[{Pizzo(1978)}]{pizzo1978}
Pizzo, V. (1978).
\newblock A three-dimensional model of corotating streams in the solar wind, 1.
  {{Theoretical}} foundations.
\newblock \emph{Journal of Geophysical Research: Space Physics} 83, 5563--5572.
\newblock \doi{10.1029/JA083iA12p05563}
\input{pizzo1978}

\bibitem[{Press et~al.(2007)}]{press2007}
Press, W., Teukolsky, S., Vetterling, W., and Flannery, B. (2007).
\newblock \emph{Numerical {{Recipes}} - {{The Art}} of {{Scientific
  Computing}}} ({Cambridge University Press}), third edn.
\input{press2007}

\bibitem[{{Price-Whelan} et~al.(2018)}]{price-whelan2018}
{Price-Whelan}, A.~M., Sip{\H o}cz, B.~M., G{\"u}nther, H.~M., Lim, P.~L.,
  Crawford, S.~M., Conseil, S., et~al. (2018).
\newblock The {{Astropy Project}}: {{Building}} an {{Open-science Project}} and
  {{Status}} of the v2.0 {{Core Package}}.
\newblock \emph{The Astronomical Journal} 156, 123.
\newblock \doi{10.3847/1538-3881/aabc4f}
\input{price-whelan2018}

\bibitem[{Reiss et~al.(2020)}]{reiss2020}
Reiss, M.~A., MacNeice, P.~J., Muglach, K., Arge, C.~N., M{\"o}stl, C., Riley,
  P., et~al. (2020).
\newblock Forecasting the {{Ambient Solar Wind}} with {{Numerical Models}}.
  {{II}}. {{An Adaptive Prediction System}} for {{Specifying Solar Wind Speed}}
  near the {{Sun}}.
\newblock \emph{The Astrophysical Journal} 891, 165.
\newblock \doi{10.3847/1538-4357/ab78a0}
\input{reiss2020}

\bibitem[{Riley et~al.(2001)}]{Riley2001}
Riley, P., Linker, J.~A., and Miki{\'c}, Z. (2001).
\newblock An empirically-driven global {{MHD}} model of the solar corona and
  inner heliosphere.
\newblock \emph{Journal of Geophysical Research: Space Physics} 106,
  15889--15901.
\newblock \doi{10.1029/2000JA000121}
\input{Riley2001}

\bibitem[{Riley and Lionello(2011)}]{riley2011}
Riley, P. and Lionello, R. (2011).
\newblock Mapping {{Solar Wind Streams}} from the {{Sun}} to 1 {{AU}}: {{A
  Comparison}} of {{Techniques}}.
\newblock \emph{Solar Physics} 270, 575--592.
\newblock \doi{10.1007/s11207-011-9766-x}
\input{riley2011}

\bibitem[{Riley et~al.(2018)}]{riley2018}
Riley, P., Mays, M.~L., Andries, J., Amerstorfer, T., Biesecker, D., Delouille,
  V., et~al. (2018).
\newblock Forecasting the {{Arrival Time}} of {{Coronal Mass Ejections}}:
  {{Analysis}} of the {{CCMC CME Scoreboard}}.
\newblock \emph{Space Weather} 16, 1245--1260.
\newblock \doi{10.1029/2018SW001962}
\input{riley2018}

\bibitem[{Robitaille et~al.(2013)}]{robitaille2013}
Robitaille, T.~P., Tollerud, E.~J., Greenfield, P., Droettboom, M., Bray, E.,
  Aldcroft, T., et~al. (2013).
\newblock Astropy: {{A}} community {{Python}} package for astronomy.
\newblock \emph{Astronomy \& Astrophysics} 558, A33.
\newblock \doi{10.1051/0004-6361/201322068}
\input{robitaille2013}

\bibitem[{Van~Leeuwen(2009)}]{vanleeuwen2009}
Van~Leeuwen, P.~J. (2009).
\newblock Particle filtering in geophysical systems.
\newblock \emph{Monthly Weather Review} 137, 4089--4114.
\newblock \doi{10.1175/2009MWR2835.1}
\input{vanleeuwen2009}

\bibitem[{Yeates et~al.(2010)}]{yeates2010}
Yeates, A.~R., Mackay, D.~H., {van Ballegooijen}, A.~A., and Constable, J.~A.
  (2010).
\newblock A nonpotential model for the {{Sun}}'s open magnetic flux.
\newblock \emph{Journal of Geophysical Research: Space Physics} 115.
\newblock \doi{10.1029/2010JA015611}
\input{yeates2010}

\end{thebibliography}
\end{document}